\documentclass[aps,preprint,prb]{revtex4}
\usepackage[T1]{fontenc}
\usepackage[latin9]{inputenc}
\usepackage[letterpaper]{geometry}
\usepackage{float}
\usepackage{amsmath}
\usepackage{graphicx}
\usepackage{amssymb}
\usepackage{pslatex}

\begin{document}

\title{Bandgap engineering of zigzag graphene nanoribbons by manipulating edge states via defective boundaries }

\author{Aihua Zhang$^1$, Yihong Wu$^3$, San-Huang Ke$^4$, Yuan Ping Feng$^1$, Chun Zhang$^{1,2}$}
\email{phyzc@nus.edu.sg}

\affiliation{
	$^1$Department of Physics, National University of Singapore, 
		2 Science Drive 3, Singapore 117542\\
	$^2$Department of Chemistry, National University of Singapore,
		3 Science Drive 3, Singapore 117543\\
	$^3$Department of Electrical and Computer Engineering, National University of Singapore, 4 Engineering Drive 3, Singapore, 117576\\
	$^4$Department of Physics, Tongji University, 1239 Siping Road, Shanghai 200092, P. R. China}

%\author{Aihua Zhang}
%\affiliation{Department of Physics, National University of Singapore, 2 Science Drive 3, Singapore, 117542}

%\author{Yihong Wu}
%\affiliation{Department of Electrical and Computer Engineering, National University of Singapore, 4 Engineering Drive 3, Singapore, 117576}

%\author{San-Huang Ke}
%\affiliation{Department of Physics, Tongji University, 1239 Siping Road, Shanghai 200092, P. R. China}

%\author{Yuan Ping Feng}
%\affiliation{Department of Physics, National University of Singapore, 2 Science Drive 3, Singapore, 117542}

%\author{Chun Zhang}
%\email{phyzc@nus.edu.sg}
%\affiliation{Department of Physics, National University of Singapore, 2 Science Drive 3, Singapore, 117542}
%\affiliation{Department of Chemistry, National University of Singapore, 3 Science Drive 3, Singapore, 117543}

\date{\today}

\begin{abstract}
One of severe limits of graphene nanoribbons (GNRs) in future applications is that zigzag GNRs (ZGNRs) are gapless, so cannot be used in field effect transistors (FETs). In this paper, using tight-binding approach and first principles method, we derived
and proved a general edge (boundary) condition for the opening of a significant bandgap in
ZGNRs with defective edge structures. The proposed semiconducting GNRs have some interesting properties including the one that they can be embedded
and integrated in a large piece of graphene without the need of completely cutting them out. We also demonstrated
a new type of high-performance all-ZGNR FET.
\end{abstract}


\maketitle

\section{Introduction}
Graphene has attracted intensive research efforts due to its unique electronic and mechanical properties.~\cite{NovoselovGMJZDGF04,NovoselovGMJKGDF05,GeimN07}
A recent experiment demonstrated a beautiful technique in fabricating graphene nanoribbons (GNRs) with atomically precise edges, suggesting the great potential of GNRs in future applications of graphene-based high performance electronics.~\cite{CaiRJBBBMSSFMF10} Theoretical calculations showed that only two thirds of
armchair graphene nanoribbons (AGNRs) with different widths are semiconducting,
while zigzag graphene nanoribbons (ZGNRs) are gapless due to localized
edge states at the Fermi level.~\cite{PhysRevB.54.17954,PhysRevLett.97.216803}
It was also theoretically demonstrated that these bandgap-closing
edge states survive in zigzag GNRs with a mixture of zigzag and armchair
sites at boundaries.~\cite{PhysRevB.54.17954,PhysRevB.80.155415} Another recent theoretical
work proved that the confinement by \emph{minimal} boundaries generally
does not produce an insulating GNR except for the armchair case.~\cite{PhysRevB.77.085423}
This theoretically predicted edge or orientation and width dependence of bandgap opening in GNRs provide serious limits in real applications of GNR-based electronic devices: First, gapless ZGNRs cannot be used in FETs, and second, the precise control of the width of AGNRs is required.  

In this paper, using the tight-binding approach and the first principles
method based on density functional theory (DFT), we derived and proved that when the number of A-site defects equals
to that of B-site defects at each boundary (A, B denotes two sublattices
of graphene), localized edge states in GNRs will be eliminated, and then a bandgap that is inversely proportional to the ribbon width will generally be open. We then showed that ZGNRs with defective boundaries that satisfy the
bandgap opening conditions can be embedded and integrated in a large
piece of graphene, which may have implications for the future design
of graphene-based integrated circuits. At last, we demonstrate a new
type of field effect transistor completely made of ZGNRS. Note that
in all previously theoretically proposed GNR-based transistors,~\cite{YanHYZZWGLD07}
the AGNR is indispensable. It is worthy mentioning here that since the long-range magnetic order is not stable in one-dimensional systems under finite temprature, we therefore stick to the non-magnetic case in this study.

\section{Results and discussion}

\begin{figure}[H]
\begin{centering}
\includegraphics{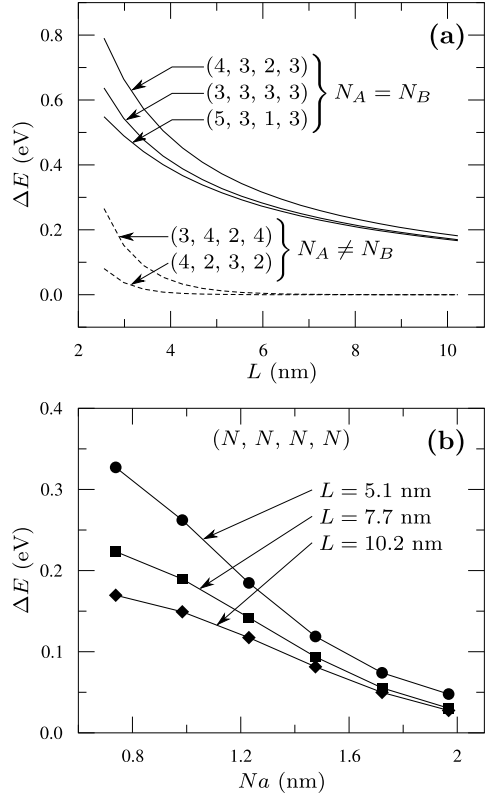} 
\par\end{centering}

\caption{\label{fig:gapvar}(a) The variation of bandgaps as a function of
the nanoribbon width ($L$) for different edge structures. (b) The
variation of bandgaps as a function of the segment length ($Na$)
of a ($N$, $N$, $N$, $N$) edge structure for different widths.}

\end{figure}

We first focus on a ZGNR with an edge structure with defects
as shown in Fig.~\ref{fig:pitted-ribbon}(a). The edge structure
can be specified by a quadruple of segment lengths in unit of the
graphene lattice constant ($a=2.46\mbox{ \AA}$), ($N_{B,1}$, $N_{A,1}$,
$N_{B,2}$, $N_{A,2}$). So the number of two-coordinated carbon atoms
at edge belong to $A$ ($B$) sublattice, $N_{A}$ ($N_{B}$), equals
to $N_{A,1}+N_{A,2}$ ($N_{B,1}+N_{B,2}$). The electronic structure
was calculated using the tight-binding approach. Only the nearest-neighbor
hopping energy (-2.7 eV) was taken into account. It is well known
that there exist edge states with $E=0$ for a semi-infinite graphene
with a zigzag edge if $2\pi/3<k_{y}a\leqslant\pi$. The edge state
entirely localizes at edge for $k_{y}a=\pi$, and otherwise decays
exponentially away from the edge.~\cite{PhysRevB.54.17954} When
two zigzag edges form a ZGNR and the edge states from both edges interact
with each other, the edge states still degenerate at $k_{y}a=\pi$,
while a small gap that decreases exponentially with the nanoribbon
width opens elsewhere. The band structure of a perfect ZGNR in a supercell
corresponding to (9, 0, 0, 0) is reproduced in Fig.~\ref{fig:pitted-ribbon}(b).
Due to band folding, there are six bands (marked as red) corresponding
to $2\pi/3<k_{y}a\leqslant\pi$. We find that these bands originating
from edge states are removed and thus an energy gap opens if $N_{A}=N_{B}$.
An example of the band structure corresponding to a (3, 3, 3, 3) edge
structure is shown in Fig.~\ref{fig:pitted-ribbon}(c), and the squared
wave functions in the inset clearly indicate they are extended states
with the form, $\sin(k_{n}x)$, which has more nodes for larger energies.
Therefore, the energy of the valence band maximum (VBM), hence the
energy gap, is inversely proportional to the nanoribbon width.
For the case of either $N_{A}<N_{B}$ or $N_{A}>N_{B}$, some of the
edge states will remain as shown in Figs.~\ref{fig:pitted-ribbon}(d)
and \ref{fig:pitted-ribbon}(e), though the perfectly localized edge
state is destroyed. The squared wave functions in the insets of Figs.~\ref{fig:pitted-ribbon}(d) and \ref{fig:pitted-ribbon}(e)
show the edge states are exponentially decaying away from the edge,
so the energy gap due to the interaction between the states at opposite
edges also decreases exponentially with respect to the nanoribbon
width. These two distinct behaviors of the bandgap variation as a
function of the nanoribbon width can be seen in Fig.~\ref{fig:gapvar}(a).
The variation of bandgaps with respect to the characteristic length
of the edge structure is shown in Fig.~\ref{fig:gapvar}(b). The
possibility to tune bandgaps with different edge structures on the
same nanoribbon might provide useful implication in the design of
nanoribbon-based electronic devices.

The above mentioned condition for the elimination of edges states that leads to bandgap opening can be understood
from the following arguments. Considering a semi-infinite graphene
sheet with $N_{A}^{e}$ ($N_{B}^{e}$) two-coordinated carbon atoms
at the edge and $N_{A}^{b}$ ($N_{B}^{b}$) three-coordinated carbon
atoms in the bulk belonging to $A$ ($B$) sublattice, we have the
following equation by the conservation of coordinate numbers,\[
2N_{A}^{e}+3N_{A}^{b}=2N_{B}^{e}+3N_{B}^{b}.\]
 Since the total number of carbon atoms in each sublattice is $N_{A(B)}^{t}=N_{A(B)}^{e}+N_{A(B)}^{b}$,
the above equation can be rewritten as $3(N_{A}^{t}-N_{B}^{t})=N_{A}^{e}-N_{B}^{e}$,
which means the relation between $N_{A}^{t}$ and $N_{B}^{t}$ is
the same as that between $N_{A}^{e}$ and $N_{B}^{e}$. The band structure
of graphene in the tight-binding approximation is calculated by\begin{align}
E\psi_{A}(\mathbf{r}) & =  t[\psi_{B}(\mathbf{r})+\psi_{B}(\mathbf{r}-\mathbf{R}_{1})+\psi_{B}(\mathbf{r}-\mathbf{R}_{2})]\label{eq:tb-b}\\
E\psi_{B}(\mathbf{r}) & =  t[\psi_{A}(\mathbf{r})+\psi_{A}(\mathbf{r}+\mathbf{R}_{1})+\psi_{A}(\mathbf{r}+\mathbf{R}_{2})],\label{eq:tb-a}\end{align}
 where $t$ is the hopping energy, $\psi_{A}(\mathbf{r})$ and $\psi_{B}(\mathbf{r})$
are the wave functions on $A$ and $B$ atoms belonging to the same
unit cell at a discrete coordinate $\mathbf{r}$, and $\mathbf{R}_{1}$
and $\mathbf{R}_{2}$ are graphene lattice vectors as shown in Fig.~\ref{fig:pitted-ribbon}(a).
For $E=0$, Eqs.~\ref{eq:tb-b} and \ref{eq:tb-a} are decoupled.
There are $N_{A}^{t}$ equations with $N_{B}^{t}$ unknowns for \ref{eq:tb-b}
and $N_{B}^{t}$ equations with $N_{A}^{t}$ unknowns for \ref{eq:tb-a}.
So if $N_{A}^{t}>N_{B}^{t}$, \ref{eq:tb-b} will have no solution
while \ref{eq:tb-a} will have solutions on $A$ sublattice.
Similar conclusion will arrive for $N_{A}^{t}<N_{B}^{t}$. The fact
that the wave function will reside on the sublattice with more atoms
can be observed by comparison of the insets in Figs.~\ref{fig:pitted-ribbon}(d)
and \ref{fig:pitted-ribbon}(e). If $N_{A}^{t}=N_{B}^{t}$ and the break of symmetry leads to no linear dependence among equations,
then Eqs.~\ref{eq:tb-b} and \ref{eq:tb-a} will have only zero solution, which is not admissible
and results in the elimination of localized states with $E=0$ that decay exponentially from the edge. 
From the above arguments, it can be seen that this boundary condition is a necessary condition for opening bandgaps in GNRs
that scale inversely with the width instead of exponentially.
Our test calculations also show that this boundary condition of bandgap opening also
applies for many other nanoribbons with general
orientations and defective edge structures.

\begin{figure}[H]
\begin{centering}
\includegraphics{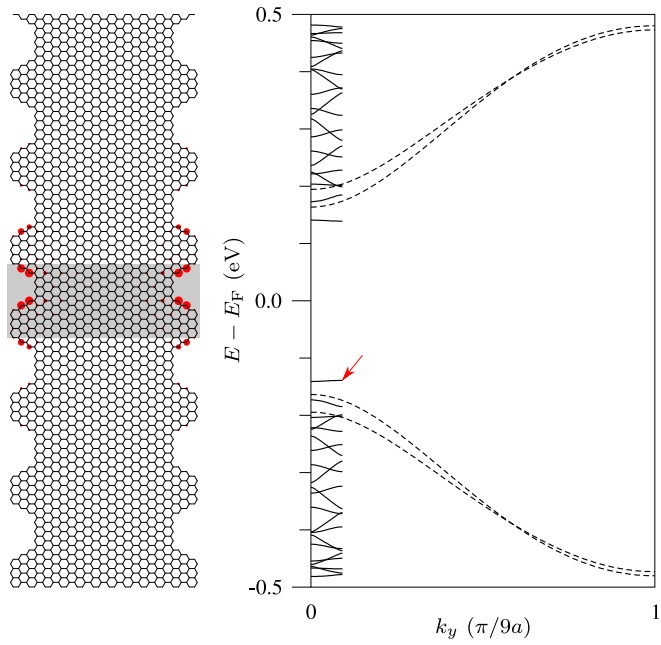} 
\par\end{centering}

\caption{\label{fig:defect}The comparison between band structures of graphene
nanoribbons with a pure (3, 3, 3, 3) edge structure (dashed line)
and with one (3, 3, 2, 3) unit (as shaded) plus ten (3, 3, 3, 3) units
in a supercell (solid line). The squared wave function (only part
of the supercell is depicted) shown in the left panel suggests that
the corresponding state (as arrowed) is localized around the defective
(3, 3, 2, 3) unit. }

\end{figure}

In practice, disorders are inevitable in the fabrication of GNRs that involves in the lithographic patterning and etching. We therefore considered a disordered
edge structure as shown in Fig.~\ref{fig:defect}: A ZGNR with one
(3, 3, 2, 3) unit (as shaded in the figure) plus 10 (3, 3, 3, 3) units
in one supercell. Here, the (3, 3, 2, 3) unit can be treated as a
disorder in the (3, 3, 3, 3) edge structure. Our tight-binding calculations
show that the disorder will induce localized states inside the bandgap
(as shown in the figure) that may not contribute to transport, and
compared to the edge structure without the disorder, the energy gap
between extended states in the disordered structure is bigger than
the bandgap of the perfect one. In real experiments, statistically
speaking, the gap-opening boundary condition is always satisfied,
so our findings presented here may have implications for recent experiments
showing that the transport gap of GNRs is inversely proportional to
the width, and independent on the orientations or edges of GNRs.~\cite{PhysRevLett.98.206805}

An interesting property of nanoribbons with a bandgap-opening edge
structure is that if a wide nanoribbon is joined with a narrow nanoribbon,
the electronic structure of the wide nanoribbon near the Fermi energy
is not altered with electrons still confined in the wide nanoribbon.
An example of a nanoribbon with a width of $L=12\sqrt{3}a$ and a
(3, 3, 3, 3) edge structure joined with a nanoribbon with the same
edge structure and a width of $L=5\sqrt{3}a$ is shown in Fig.~\ref{fig:ribconfine}(a).
The band structure of the compound system near the Fermi energy (the
conduction and valence band) in Fig.~\ref{fig:ribconfine}(b) is
almost the same as that of the stand-alone nanoribbon shown in Fig.~\ref{fig:pitted-ribbon}(c).
The charge distribution of the state at VBM in Fig.~\ref{fig:ribconfine}(a)
and the local density of states in Fig.~\ref{fig:ribconfine}(c)
clearly indicate that the wave function is only localized in the wide
nanoribbon. The confinement can be understood from the bandgap difference
of two nanoribbons with different widths. Note that the integrated GNRs discussed here can be also regarded as a special type of graphene antidot lattice structures proposed earlier.~\cite{PhysRevLett.100.136804, ZhangTDFZ11} This property makes it possible
to fabricate individual nanoribbon-based electronic devices by patterning
rows of holes in a large piece of graphene avoiding complete cutout
and glued together, which might be beneficial for the integration
of future graphene-based electric circuits. On experimental side, the patterned graphene nanostructures discussed here can be obtained experimentally by using techniques such as templated self-assembly of block copolymers~\cite{ChengRST06} or direct writing using a helium ion beam.~\cite{LemmeBWSBJM09} The periodically patterned structure may be formed by first forming resist patterns on the graphene sheet followed by templated self-assembly of block copolymers in the region where the resist have been removed and etching of graphene by using the copolymer patterns as the mask. On the other hand, the random patterns can be formed by direct writing using a helium ion beam. Prior to the lithography processes, an alignment mark may be formed on the wafer by using an appropriate graphene edge as a reference so as to align the patterns in specific directions with respect to the underlying graphene lattice structure.

\begin{figure}[H]
\begin{centering}
\includegraphics{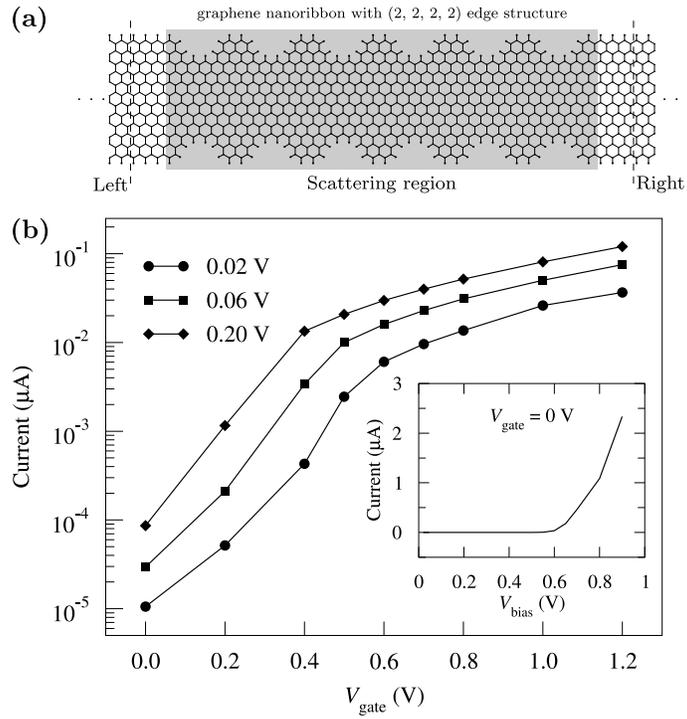} 
\par\end{centering}

\caption{\label{fig:ivcurve}(a) The configuration for the transport calculation.
(b) The variation of current as a function of gate voltage for different
bias voltages. The inset shows the bias voltage dependence of current
for zero gate voltage.}

\end{figure}

At last, we show a FET completely made of ZGNRs as shown in Fig.~\ref{fig:ivcurve}.
In all previous theoretically proposed GNR-based FETs, the AGNR is
an indispensable component due to the fact that pure ZGNR is metallic.
Here, the proposed FET consists of two pure ZGNR electrodes (left
and right), and a ZGNR with a defective (2, 2, 2, 2) edge structure.
The transport calculations were done using a first principles approach
combining the non-equilibrium Green's function's techniques and DFT
via the ATK code.~\cite{PhysRevB.63.245407,PhysRevB.65.165401,SolerAGGJOS02}
In the inset of the figure, the current-voltage (I-V) curve is shown
for the zero gate voltage. The bias range of the zero current comes from the bandgap of the defective ZGNR in the center, confirming the bandgap opening condition we derived from the tight binding approach. The currents as a function of gate voltage for different bias
voltages suggest that the on-off ratio of this proposed FET is bigger
than 1000. Compared to the previously proposed all-GNR based FET that
used two ZGNRs and one AGNR,~\cite{YanHYZZWGLD07} the FET suggested here has two obvious advantages:
First, the complicated contacts between
differently orientated AGNR and ZGNRs are avoided. Second, the precise control of ribbon width is not required. 

\section{Conclusion}
In conclusion, using the tight-binding approach, we derived a general boundary condition for the band gap opening in the ZGNRs with defective edges: When the number of A-site defects equals to that of B-site defects, the ZGNRs are semiconducting. We further showed that the semiconducting ZGNRs generated this way can be integrated in a large piece of graphene by correctly patterning holes, which may be useful for the future large-scale integration of GNR-based devices. At last, we demonstrated using first principles calculations a high-performance FET completely made of ZGNRs. Results presented in this paper may be used to explain the recent experimental measurements showing that the transport gap always exists independent of the crystallographic orientations of GNRs. We expect these findings to provide impetus for new experiments as well motivations for new ideas in designing ZGNR-based electronic devices.    

\section{Acknowledgment}
We thank Professor A. H. Castro Neto and Dr. V. M. Pereira for stimulating and helpful discussions. This work was supported by NUS Academic Research Fund (Grant Nos: R-144-000-237133 and R-144-000-255-112). Computations were performed
at the Centre for Computational Science and Engineering at NUS.

\newpage{}

\begin{center}
\includegraphics[width=3.4in]{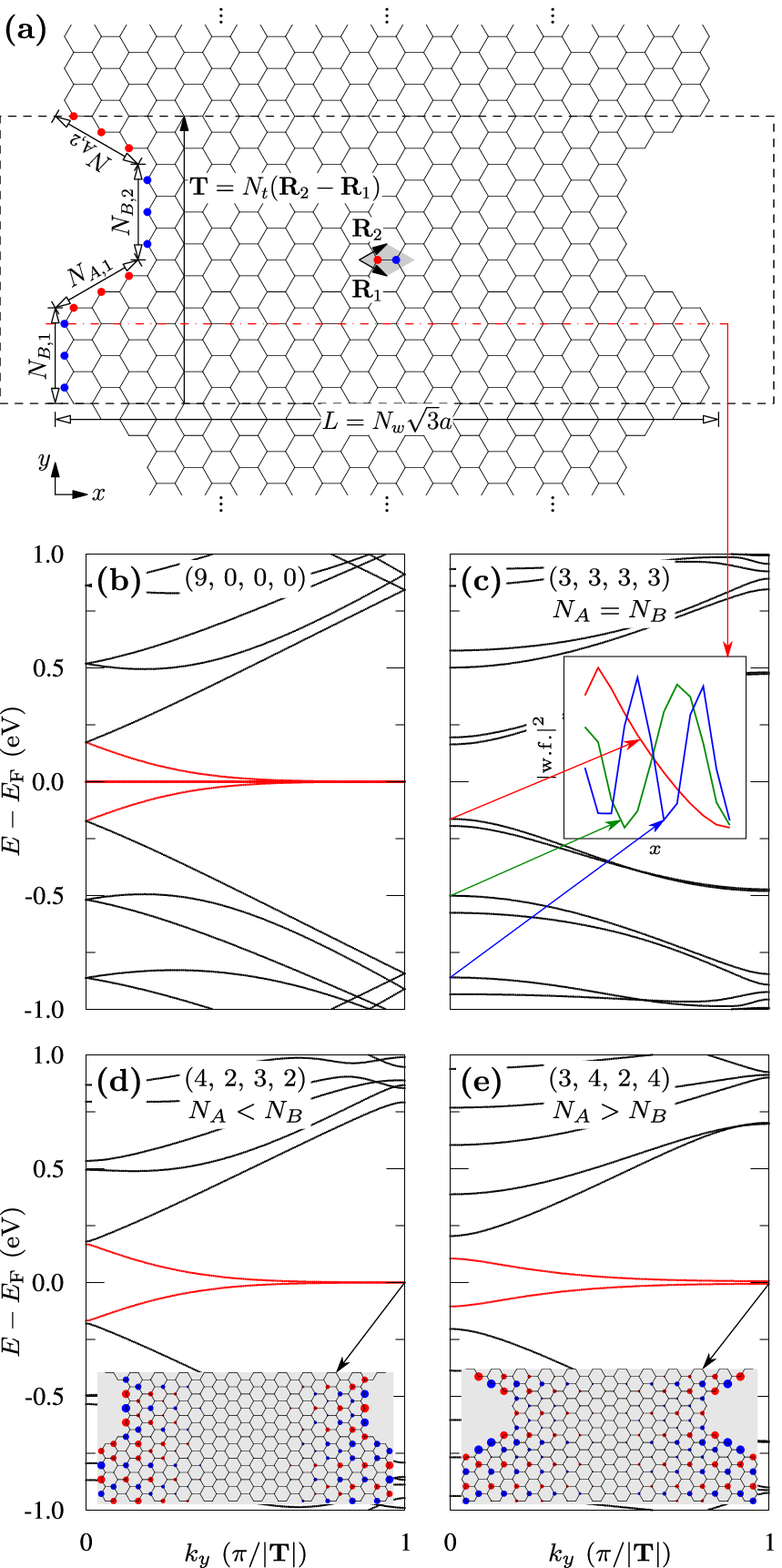} 
\par\end{center}

\begin{figure}[H]
\caption{\label{fig:pitted-ribbon}(a) The lattice structure of a zigzag graphene
nanoribbon with periodic edge structures. The unit cell is indicated
by the dashed line. The edge structure is denoted by a quadruple of
($N_{B,1}$, $N_{A,1}$, $N_{B,2}$, $N_{A,2}$), each number of which
corresponds to the segment length in unit of the graphene lattice
constant, $a$. Other parameters of the system are the nanoribbon
width ($L$) and the translational vector ($\mathbf{T}$). Carbon
atoms belonging to different sublattices at edge are designated red
($A$) and blue ($B$) colors. (b-e) The band structures of nanoribbons
with different edge structures and the same width $L=12\sqrt{3}a$.
In the inset of (c) are shown the squared wave functions along the
dash-dotted line in (a) for different states at $k_{y}=0$. The squared
wave functions corresponding to the valence band maximums are also
plotted in the insets of (d) and (e). The radii of filled discs are
proportional to $R(\log_{10}|\psi(\mathbf{r})|^{2}+4)$, where $R(x)$
is the ramp function, and the color is determined by the sign of real
part of $\psi(\mathbf{r})$.}

\end{figure}

\begin{figure}[H]
\begin{centering}
\includegraphics[width=3.4in]{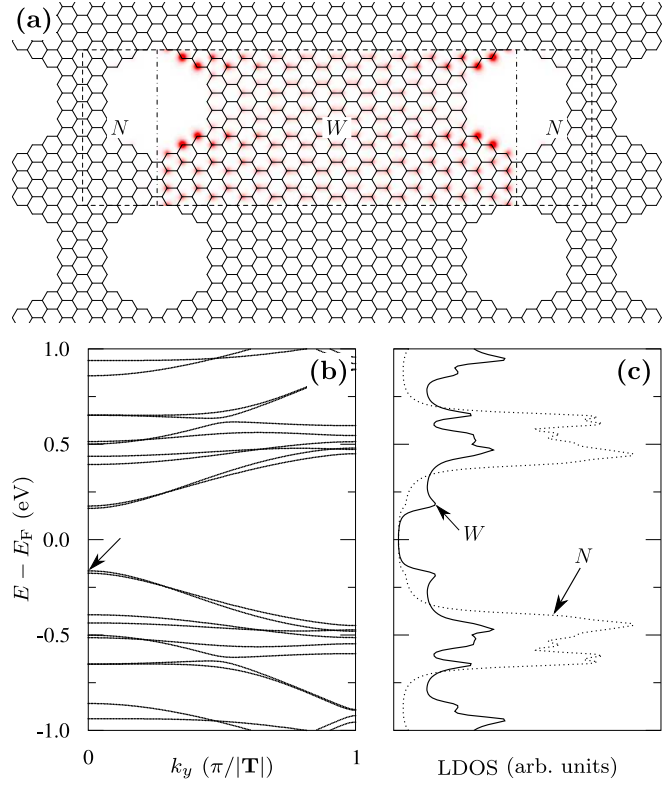} 
\par\end{centering}

\caption{\label{fig:ribconfine}(a) The unit cell (dashed line) of a wide nanoribbon
($W$) joined with a narrow nanoribbon ($N$) having the same edge
structure. The nanoribbon $W$ between two dash-dotted lines is identical
to the nanoribbon shown in Fig.~\ref{fig:pitted-ribbon}(a). (b)
The band structure of the system in (a). The squared wave function
plotted in (a) corresponds to the state indicated by an arrow. (c)
The corresponding local density of states in $W$ and $N$ as shown
in (a).}

\end{figure}

\bibliography{ref}

\begin{thebibliography}{17}
\expandafter\ifx\csname natexlab\endcsname\relax\def\natexlab#1{#1}\fi
\expandafter\ifx\csname bibnamefont\endcsname\relax
  \def\bibnamefont#1{#1}\fi
\expandafter\ifx\csname bibfnamefont\endcsname\relax
  \def\bibfnamefont#1{#1}\fi
\expandafter\ifx\csname citenamefont\endcsname\relax
  \def\citenamefont#1{#1}\fi
\expandafter\ifx\csname url\endcsname\relax
  \def\url#1{\texttt{#1}}\fi
\expandafter\ifx\csname urlprefix\endcsname\relax\def\urlprefix{URL }\fi
\providecommand{\bibinfo}[2]{#2}
\providecommand{\eprint}[2][]{\url{#2}}

\bibitem[{\citenamefont{Novoselov et~al.}(2004)\citenamefont{Novoselov, Geim,
  Morozov, Jiang, Zhang, Dubonos, Grigorieva, and Firsov}}]{NovoselovGMJZDGF04}
\bibinfo{author}{\bibfnamefont{K.~S.} \bibnamefont{Novoselov}},
  \bibinfo{author}{\bibfnamefont{A.~K.} \bibnamefont{Geim}},
  \bibinfo{author}{\bibfnamefont{S.~V.} \bibnamefont{Morozov}},
  \bibinfo{author}{\bibfnamefont{D.}~\bibnamefont{Jiang}},
  \bibinfo{author}{\bibfnamefont{Y.}~\bibnamefont{Zhang}},
  \bibinfo{author}{\bibfnamefont{S.~V.} \bibnamefont{Dubonos}},
  \bibinfo{author}{\bibfnamefont{I.~V.} \bibnamefont{Grigorieva}},
  \bibnamefont{and} \bibinfo{author}{\bibfnamefont{A.~A.}
  \bibnamefont{Firsov}}, \bibinfo{journal}{Science}
  \textbf{\bibinfo{volume}{306}}, \bibinfo{pages}{666} (\bibinfo{year}{2004}).

\bibitem[{\citenamefont{Novoselov et~al.}(2005)\citenamefont{Novoselov, Geim,
  Morozov, Jiang, Katsnelson, Grigorieva, Dubonos, and
  Firsov}}]{NovoselovGMJKGDF05}
\bibinfo{author}{\bibfnamefont{K.~S.} \bibnamefont{Novoselov}},
  \bibinfo{author}{\bibfnamefont{A.~K.} \bibnamefont{Geim}},
  \bibinfo{author}{\bibfnamefont{S.~V.} \bibnamefont{Morozov}},
  \bibinfo{author}{\bibfnamefont{D.}~\bibnamefont{Jiang}},
  \bibinfo{author}{\bibfnamefont{M.~I.} \bibnamefont{Katsnelson}},
  \bibinfo{author}{\bibfnamefont{I.~V.} \bibnamefont{Grigorieva}},
  \bibinfo{author}{\bibfnamefont{S.~V.} \bibnamefont{Dubonos}},
  \bibnamefont{and} \bibinfo{author}{\bibfnamefont{A.~A.}
  \bibnamefont{Firsov}}, \bibinfo{journal}{Nature}
  \textbf{\bibinfo{volume}{438}}, \bibinfo{pages}{197} (\bibinfo{year}{2005}).

\bibitem[{\citenamefont{Geim and Novoselov}(2007)}]{GeimN07}
\bibinfo{author}{\bibfnamefont{A.~K.} \bibnamefont{Geim}} \bibnamefont{and}
  \bibinfo{author}{\bibfnamefont{K.~S.} \bibnamefont{Novoselov}},
  \bibinfo{journal}{Nat. Mater.} \textbf{\bibinfo{volume}{6}},
  \bibinfo{pages}{183} (\bibinfo{year}{2007}).

\bibitem[{\citenamefont{Cai et~al.}(2010)\citenamefont{Cai, Ruffieux, Jaafar,
  Bieri, Braun, Blankenburg, Muoth, Seitsonen, Saleh, Feng
  et~al.}}]{CaiRJBBBMSSFMF10}
\bibinfo{author}{\bibfnamefont{J.~M.} \bibnamefont{Cai}},
  \bibinfo{author}{\bibfnamefont{P.}~\bibnamefont{Ruffieux}},
  \bibinfo{author}{\bibfnamefont{R.}~\bibnamefont{Jaafar}},
  \bibinfo{author}{\bibfnamefont{M.}~\bibnamefont{Bieri}},
  \bibinfo{author}{\bibfnamefont{T.}~\bibnamefont{Braun}},
  \bibinfo{author}{\bibfnamefont{S.}~\bibnamefont{Blankenburg}},
  \bibinfo{author}{\bibfnamefont{M.}~\bibnamefont{Muoth}},
  \bibinfo{author}{\bibfnamefont{A.~P.} \bibnamefont{Seitsonen}},
  \bibinfo{author}{\bibfnamefont{M.}~\bibnamefont{Saleh}},
  \bibinfo{author}{\bibfnamefont{X.~L.} \bibnamefont{Feng}},
  \bibnamefont{et~al.}, \bibinfo{journal}{Nature}
  \textbf{\bibinfo{volume}{466}}, \bibinfo{pages}{470} (\bibinfo{year}{2010}).

\bibitem[{\citenamefont{Nakada et~al.}(1996)\citenamefont{Nakada, Fujita,
  Dresselhaus, and Dresselhaus}}]{PhysRevB.54.17954}
\bibinfo{author}{\bibfnamefont{K.}~\bibnamefont{Nakada}},
  \bibinfo{author}{\bibfnamefont{M.}~\bibnamefont{Fujita}},
  \bibinfo{author}{\bibfnamefont{G.}~\bibnamefont{Dresselhaus}},
  \bibnamefont{and} \bibinfo{author}{\bibfnamefont{M.~S.}
  \bibnamefont{Dresselhaus}}, \bibinfo{journal}{Phys. Rev. B}
  \textbf{\bibinfo{volume}{54}}, \bibinfo{pages}{17954} (\bibinfo{year}{1996}).

\bibitem[{\citenamefont{Son et~al.}(2006)\citenamefont{Son, Cohen, and
  Louie}}]{PhysRevLett.97.216803}
\bibinfo{author}{\bibfnamefont{Y.-W.} \bibnamefont{Son}},
  \bibinfo{author}{\bibfnamefont{M.~L.} \bibnamefont{Cohen}}, \bibnamefont{and}
  \bibinfo{author}{\bibfnamefont{S.~G.} \bibnamefont{Louie}},
  \bibinfo{journal}{Phys. Rev. Lett.} \textbf{\bibinfo{volume}{97}},
  \bibinfo{pages}{216803} (\bibinfo{year}{2006}).

\bibitem[{\citenamefont{Ihnatsenka et~al.}(2009)\citenamefont{Ihnatsenka,
  Zozoulenko, and Kirczenow}}]{PhysRevB.80.155415}
\bibinfo{author}{\bibfnamefont{S.}~\bibnamefont{Ihnatsenka}},
  \bibinfo{author}{\bibfnamefont{I.~V.} \bibnamefont{Zozoulenko}},
  \bibnamefont{and}
  \bibinfo{author}{\bibfnamefont{G.}~\bibnamefont{Kirczenow}},
  \bibinfo{journal}{Phys. Rev. B} \textbf{\bibinfo{volume}{80}},
  \bibinfo{pages}{155415} (\bibinfo{year}{2009}).

\bibitem[{\citenamefont{Akhmerov and Beenakker}(2008)}]{PhysRevB.77.085423}
\bibinfo{author}{\bibfnamefont{A.~R.} \bibnamefont{Akhmerov}} \bibnamefont{and}
  \bibinfo{author}{\bibfnamefont{C.~W.~J.} \bibnamefont{Beenakker}},
  \bibinfo{journal}{Phys. Rev. B} \textbf{\bibinfo{volume}{77}},
  \bibinfo{pages}{085423} (\bibinfo{year}{2008}).

\bibitem[{\citenamefont{Yan et~al.}(2007)\citenamefont{Yan, Huang, Yu, Zheng,
  Zang, Wu, Gu, Liu, and Duan}}]{YanHYZZWGLD07}
\bibinfo{author}{\bibfnamefont{Q.~M.} \bibnamefont{Yan}},
  \bibinfo{author}{\bibfnamefont{B.}~\bibnamefont{Huang}},
  \bibinfo{author}{\bibfnamefont{J.}~\bibnamefont{Yu}},
  \bibinfo{author}{\bibfnamefont{F.~W.} \bibnamefont{Zheng}},
  \bibinfo{author}{\bibfnamefont{J.}~\bibnamefont{Zang}},
  \bibinfo{author}{\bibfnamefont{J.}~\bibnamefont{Wu}},
  \bibinfo{author}{\bibfnamefont{B.~L.} \bibnamefont{Gu}},
  \bibinfo{author}{\bibfnamefont{F.}~\bibnamefont{Liu}}, \bibnamefont{and}
  \bibinfo{author}{\bibfnamefont{W.~H.} \bibnamefont{Duan}},
  \bibinfo{journal}{Nano Lett.} \textbf{\bibinfo{volume}{7}},
  \bibinfo{pages}{1469} (\bibinfo{year}{2007}).

\bibitem[{\citenamefont{Han et~al.}(2007)\citenamefont{Han, \"Ozyilmaz, Zhang,
  and Kim}}]{PhysRevLett.98.206805}
\bibinfo{author}{\bibfnamefont{M.~Y.} \bibnamefont{Han}},
  \bibinfo{author}{\bibfnamefont{B.}~\bibnamefont{\"Ozyilmaz}},
  \bibinfo{author}{\bibfnamefont{Y.}~\bibnamefont{Zhang}}, \bibnamefont{and}
  \bibinfo{author}{\bibfnamefont{P.}~\bibnamefont{Kim}},
  \bibinfo{journal}{Phys. Rev. Lett.} \textbf{\bibinfo{volume}{98}},
  \bibinfo{pages}{206805} (\bibinfo{year}{2007}).

\bibitem[{\citenamefont{Pedersen et~al.}(2008)\citenamefont{Pedersen, Flindt,
  Pedersen, Mortensen, Jauho, and Pedersen}}]{PhysRevLett.100.136804}
\bibinfo{author}{\bibfnamefont{T.~G.} \bibnamefont{Pedersen}},
  \bibinfo{author}{\bibfnamefont{C.}~\bibnamefont{Flindt}},
  \bibinfo{author}{\bibfnamefont{J.}~\bibnamefont{Pedersen}},
  \bibinfo{author}{\bibfnamefont{N.~A.} \bibnamefont{Mortensen}},
  \bibinfo{author}{\bibfnamefont{A.-P.} \bibnamefont{Jauho}}, \bibnamefont{and}
  \bibinfo{author}{\bibfnamefont{K.}~\bibnamefont{Pedersen}},
  \bibinfo{journal}{Phys. Rev. Lett.} \textbf{\bibinfo{volume}{100}},
  \bibinfo{pages}{136804} (\bibinfo{year}{2008}).

\bibitem[{\citenamefont{Zhang et~al.}(2011)\citenamefont{Zhang, Teoh, Dai,
  Feng, and Zhang}}]{ZhangTDFZ11}
\bibinfo{author}{\bibfnamefont{A.~H.} \bibnamefont{Zhang}},
  \bibinfo{author}{\bibfnamefont{H.~F.} \bibnamefont{Teoh}},
  \bibinfo{author}{\bibfnamefont{Z.~X.} \bibnamefont{Dai}},
  \bibinfo{author}{\bibfnamefont{Y.~P.} \bibnamefont{Feng}}, \bibnamefont{and}
  \bibinfo{author}{\bibfnamefont{C.}~\bibnamefont{Zhang}},
  \bibinfo{journal}{Appl. Phys. Lett.} \textbf{\bibinfo{volume}{98}},
  \bibinfo{pages}{023105} (\bibinfo{year}{2011}).

\bibitem[{\citenamefont{Cheng et~al.}(2006)\citenamefont{Cheng, Ross, Smith,
  and Thomas}}]{ChengRST06}
\bibinfo{author}{\bibfnamefont{J.~Y.} \bibnamefont{Cheng}},
  \bibinfo{author}{\bibfnamefont{C.~A.} \bibnamefont{Ross}},
  \bibinfo{author}{\bibfnamefont{H.~I.} \bibnamefont{Smith}}, \bibnamefont{and}
  \bibinfo{author}{\bibfnamefont{E.~L.} \bibnamefont{Thomas}},
  \bibinfo{journal}{Adv. Mater.} \textbf{\bibinfo{volume}{18}},
  \bibinfo{pages}{2505} (\bibinfo{year}{2006}).

\bibitem[{\citenamefont{Lemme et~al.}(2009)\citenamefont{Lemme, Bell, Williams,
  Stern, Baugher, Jarillo-Herrero, and Marcus}}]{LemmeBWSBJM09}
\bibinfo{author}{\bibfnamefont{M.~C.} \bibnamefont{Lemme}},
  \bibinfo{author}{\bibfnamefont{D.~C.} \bibnamefont{Bell}},
  \bibinfo{author}{\bibfnamefont{J.~R.} \bibnamefont{Williams}},
  \bibinfo{author}{\bibfnamefont{L.~A.} \bibnamefont{Stern}},
  \bibinfo{author}{\bibfnamefont{B.~W.~H.} \bibnamefont{Baugher}},
  \bibinfo{author}{\bibfnamefont{P.}~\bibnamefont{Jarillo-Herrero}},
  \bibnamefont{and} \bibinfo{author}{\bibfnamefont{C.~M.}
  \bibnamefont{Marcus}}, \bibinfo{journal}{ACS Nano}
  \textbf{\bibinfo{volume}{3}}, \bibinfo{pages}{2674} (\bibinfo{year}{2009}).

\bibitem[{\citenamefont{Taylor et~al.}(2001)\citenamefont{Taylor, Guo, and
  Wang}}]{PhysRevB.63.245407}
\bibinfo{author}{\bibfnamefont{J.}~\bibnamefont{Taylor}},
  \bibinfo{author}{\bibfnamefont{H.}~\bibnamefont{Guo}}, \bibnamefont{and}
  \bibinfo{author}{\bibfnamefont{J.}~\bibnamefont{Wang}},
  \bibinfo{journal}{Phys. Rev. B} \textbf{\bibinfo{volume}{63}},
  \bibinfo{pages}{245407} (\bibinfo{year}{2001}).

\bibitem[{\citenamefont{Brandbyge et~al.}(2002)\citenamefont{Brandbyge, Mozos,
  Ordej\'on, Taylor, and Stokbro}}]{PhysRevB.65.165401}
\bibinfo{author}{\bibfnamefont{M.}~\bibnamefont{Brandbyge}},
  \bibinfo{author}{\bibfnamefont{J.-L.} \bibnamefont{Mozos}},
  \bibinfo{author}{\bibfnamefont{P.}~\bibnamefont{Ordej\'on}},
  \bibinfo{author}{\bibfnamefont{J.}~\bibnamefont{Taylor}}, \bibnamefont{and}
  \bibinfo{author}{\bibfnamefont{K.}~\bibnamefont{Stokbro}},
  \bibinfo{journal}{Phys. Rev. B} \textbf{\bibinfo{volume}{65}},
  \bibinfo{pages}{165401} (\bibinfo{year}{2002}).

\bibitem[{\citenamefont{Soler et~al.}(2002)\citenamefont{Soler, Artacho, Gale,
  Garc\'ia, Junquera, Ordej\'on, and S\'anchez-Portal}}]{SolerAGGJOS02}
\bibinfo{author}{\bibfnamefont{J.~M.} \bibnamefont{Soler}},
  \bibinfo{author}{\bibfnamefont{E.}~\bibnamefont{Artacho}},
  \bibinfo{author}{\bibfnamefont{J.~D.} \bibnamefont{Gale}},
  \bibinfo{author}{\bibfnamefont{A.}~\bibnamefont{Garc\'ia}},
  \bibinfo{author}{\bibfnamefont{J.}~\bibnamefont{Junquera}},
  \bibinfo{author}{\bibfnamefont{P.}~\bibnamefont{Ordej\'on}},
  \bibnamefont{and}
  \bibinfo{author}{\bibfnamefont{D.}~\bibnamefont{S\'anchez-Portal}},
  \bibinfo{journal}{J. Phys.-Condes. Matter} \textbf{\bibinfo{volume}{14}},
  \bibinfo{pages}{2745} (\bibinfo{year}{2002}).

\end{thebibliography}

\end{document}